\documentstyle[12pt,epsf]{article}
\date{1 September 1993}

\begin{document}
\begin{center}
{\bf RECURSIVE PROPORTIONAL-FEEDBACK AND ITS USE TO\\
CONTROL CHAOS IN AN ELECTROCHEMICAL SYSTEM}\footnote{Published
in ``Proceedings of the 2nd Conference on EXPERIMENTAL CHAOS''
edited by W.\ Ditto, L.\ Pecora, S.\ Vohra, M.\ Shlesinger,
and M.\ Spano (World Scientific, River Ridge NJ, 1995) 
pp 304--316.}

\vspace{18mm}

P.\ PARMANANDA, R.\ W.\ ROLLINS, P.\ SHERARD\\%\footnote{Present 
%address: School of Physics, Georga Institute of Technology,
%Atlanta, GA~~30332}\\ 
{\em Condensed Matter and Surface Sciences Program, \\
Department of Physics and Astronomy, 
Ohio University, Athens, Ohio 45701-2979} 

\vspace{6mm}

and 

\vspace{6mm}

H.\ D.\ DEWALD\\
{\em Condensed Matter and Surface Sciences Program, \\
Department of Chemistry, Ohio University, Athens, Ohio 45701-2979} 
 
\end{center}

\vspace{8mm}
%\begin{abstract}
\begin{center}

{\small ABSTRACT}\\

\vspace{10pt}

\begin{minipage}{30pc}
\parindent=3ex
{\small We report the successful application of a simple recursive 
proportional-feedback (RPF) algorithm to
control chemical chaos observed during the electrodissolution of a 
rotating copper disk in a sodium acetate/acetic acid buffer.
The RPF method is generally applicable to the control of 
chaotic systems that are highly dissipative and adequately described by
one-dimensional maps of a single variable.
The successive minima of the measured
anodic current generate a return map that is used to implement the control
strategy and the anodic potential is the control parameter.  Unstable 
periodic orbits of period one and period two are stabilized by applying small
perturbations ($\approx$ 0.05\%) to the anodic potential on each cycle of 
the periodic oscillation. Experimental evidence is presented to indicate
why the RPF method was necessary in this system and the theoretical 
robustness of the algorithm is discussed.}
\end{minipage}
\end{center}     
%\end{abstract}
%\pacs{PACS numbers: 05.45.+b, 87.10.+e}
%\narrowtext
\vspace{0.25in}
\parindent=6ex

\noindent
{\bf 1.~~Introduction}
\vspace{14pt}

   Two experiments were recently reported describing the control of
autonomous oscillatory chemical systems in the chaotic regime.
Petrov, G\'{a}sp\'{a}r, Masere, and Showalter$^{\ref{bib:BZctrl1}}$ 
stabilized periodic orbits
in the Belousov-Zhabotinsky reaction and we recently 
reported$^{\ref{bib:RPFexp1}}$ 
the first experiment to control chaotic oscillations 
in an electrochemical cell.  In controlling the electrochemical
cell we found it necessary, in general, to use the recursive 
proportional-feedback (RPF) algorithm recently developed by 
Rollins, Parmananda, and Sherard$^{\ref{bib:RPFth1}}$.  
The RPF algorithm is simple to apply and is generally applicable to 
highly dissipative systems that are well described by a one-dimensional
Poincar\'e return map of a single measured variable.
In this paper we report the further use of the RPF algorithm to stabilize 
both period-1 and period-2 orbits in the electrochemical system.  
We also present experimental evidence that suggests why we found it
necessary to use the new RPF method 
instead of the simple proportional feedback method suggested by 
Petrov, Peng, and Showalter$^{\ref{bib:PPS2}}$ and by Hunt$^{\ref{bib:EH}}$ 
that often
works for systems well described by a one-dimensional return map.  
Finally, we briefly discuss the robustness of the RPF method.  This
discussion is based on an analysis of the RPF method in terms more akin to
standard control theory$^{\ref{bib:Ogata}}$ as described recently for the  
general high dimensional system by Romeiras, Grebogi, Ott, and 
Dayawansa$^{\ref{bib:RGOD}}$.  This approach treats the experimental system
(described by a one-dimensional return map of a single variable) together 
with a linear recursive control stratagy applied to a single control 
parameter as a discrete two-dimensional system.  The desired periodic orbit 
is a fixed point of the two-dimensional map.
The range of acceptable control conditions is determined by the requirement
that the fixed point remain stable.  The RPF control 
stratagy$^{\ref{bib:RPFth1}}$ makes the fixed point superstable.  

\vspace{14pt}
\noindent
{\bf 2.~~Experimental System}
\vspace{14pt}

\begin{figure}[hbt]
\begin{center}
%\special{landscape}
\leavevmode
\epsfxsize=4.5in
\epsffile{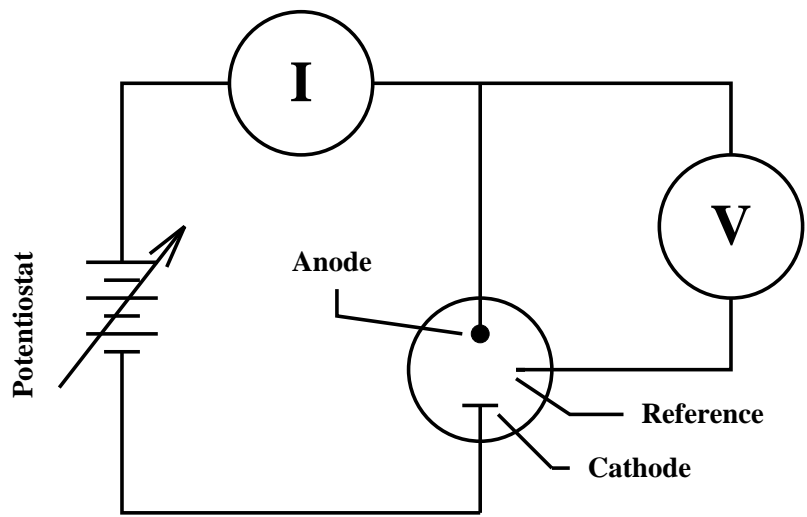}
\end{center}
\begin{center}
\begin{minipage}{30pc}
\caption[]{{\small Schematic representation of the three-electrode
electrochemical cell. $V$ is the anodic potential and $I$ is
the anodic current.  The potentiostat adjusts the emf to hold
$V$ at the desired set value.}\label{schematic}}
\end{minipage}
\end{center}

\end{figure}

The experimental system was a three-electrode 
electrochemical cell set up to study the 
potentiostatic electrodissolution of copper in an acetate
buffer electrolyte solution of sodium acetate and
acetic acid.  The experimental setup and the general electrochemical
behavior of the system is described in some detail by Dewald, Parmananda, and
Rollins$^{\ref{bib:DPR1},\ref{bib:DPR2}}$.  
Figure~\ref{schematic} is a schematic diagram
of the system.
%\vspace{3.5in}

The anode was 
a rotating copper 
disk (5~mm diameter) shrouded in a cylinder of Teflon. 
The  supporting electrolyte solution was a 
mixture of 60 parts glacial acetic acid
to 35 parts 2M sodium acetate. 
The potential of the anode was measured relative to a 
saturated calomel reference electrode (SCE) and the cathode was a  
2.5 cm$^{2}$ platinum foil disk.  The emf of the circuit was
continuously adjusted by a potentiostat to maintain the desired 
set value of the anodic potential (the potential between the anode 
and the reference electrode).  The anodic potential was used 
as our control parameter.  Time series data was collected by measuring
the anodic current at 20~ms intervals using a
12 bit A/D converter interfaced
to a computer.  The period for the current oscillations in the
chaotic states used in these experiments ranged from 2 to 4 sec so that
the time between data points in the time series is small compared 
to the typical period
of the oscillations.  The set point for the anodic potential was controlled
by the computer through a D/A converter with the smallest increment of the
control voltage being 0.1~mV.

\pagebreak
\vspace{14pt}
\noindent
{\bf 3.~~Control Method}
\vspace{14pt}

   The RPF control strategy$^{\ref{bib:RPFth1}}$ was used to stabilize both
period-1 and period-2 orbits within chaotic attractors measured
by a time series of the anodic current.  A Poincar\'{e} section was taken
at the time when the anodic current goes through a minimum.  A one-dimensional
return map was constructed from the sequence of current minima where 
$I_n$ is the current minimum at the beginning of the $n$th 
Poincar\'{e} cycle. The
value of the anodic potential during the $n$th cycle is 
$V_n = V + \delta V_n$.  According to the RPF algorithm, 
control is established by adding an increment to the anodic potential 
during the $n$th cycle given by$^{\ref{bib:RPFexp1},\ref{bib:RPFth1}}$
\begin{eqnarray}
\delta  V_n =  K( I_n - I_F) + R\,\delta V_{n-1},
\label{eq:rpf}
\end{eqnarray}
where  $I_F$ is the unstable fixed point of the target orbit 
for the return map obtained for a time series taken 
at $ V =V_0$. The value of $I_F$ and the constants $K$ and $R$, 
are determined from a precontrol
experimental procedure carried out using an interactive computer 
program for data acquisition and display.  
  
   The precontrol procedure is described in detail in 
references$^{\ref{bib:RPFexp1},\ref{bib:RPFth1}}$.  Firstly, the 
fixed point $I_F$ and
the slope, $\mu$, of the one-dimensional return map was determined by
a least square fitting of the 
displayed return map ($I_n$ versus $I_{n-1}$) for a sequence of 
current minima collected in the neighborhood of the desired fixed point
with the anodic potential
held constant at $V = V_0$.  Secondly, a sequence of current minima were
collected and displayed as a return map while the anodic potential
is changed back and forth between two values; $V_n = V_0$ for $n$ 
odd and $V_n = V_0 + \Delta V$ for $n$ even.  The value of $\Delta V > 0$
was chosen to be about the size of the maximum $\delta V_n$ 
used later during control.  If the system is sufficiently dissipative, 
as our electrochemical system was, the resulting return map
consists of two curves called the {\em up} and {\em back} maps formed from 
alternate ($I_{n-1}$, $I_n$) pairs with $n$ odd and even respectively.
The shift in the fixed point per unit $\Delta V$ is measured by least squares
fitting for 
the {\em up} and {\em back} maps
giving $g_u$ and $g_b$ respectively.   The values of $K$ and $R$ were then 
calculated using the RPF relations$^{\ref{bib:RPFth1}}$    
\begin{equation}
 K = \frac{ \mu^2}{( \mu -1)( \mu  g_u + g_b)}, \qquad
 R =  \frac{ -\mu  g_b}{( \mu  g_u + g_b)}.
\label{eq:KR2}
\end{equation}
Typically these control constants, $K$ and $R$, could be 
determined within 10--20 minutes (corresponding to approximately 
200 cycles of the experimental 
return map) from the start of data acquisition. 

With these control constants available, the control algorithm was
initiated. The anodic potential was changed according to the RPF 
algorithm of Eq.~\ref{eq:rpf} whenever a minimum of the anodic
current, $I_n$, came within about 0.2 mA of the measured
fixed point $I_F$.
 
\vspace{14pt}
\noindent
{\bf 4.~~Results and Discussion}
\vspace{14pt}

\noindent
{\em 4.1.~~Control of period-1 and period-2 orbits}
\vspace{14pt}

\begin{figure}[hbt]
\begin{center}
%\vspace{7.0 in}
\leavevmode
\epsfxsize=4.3in
\epsffile{ 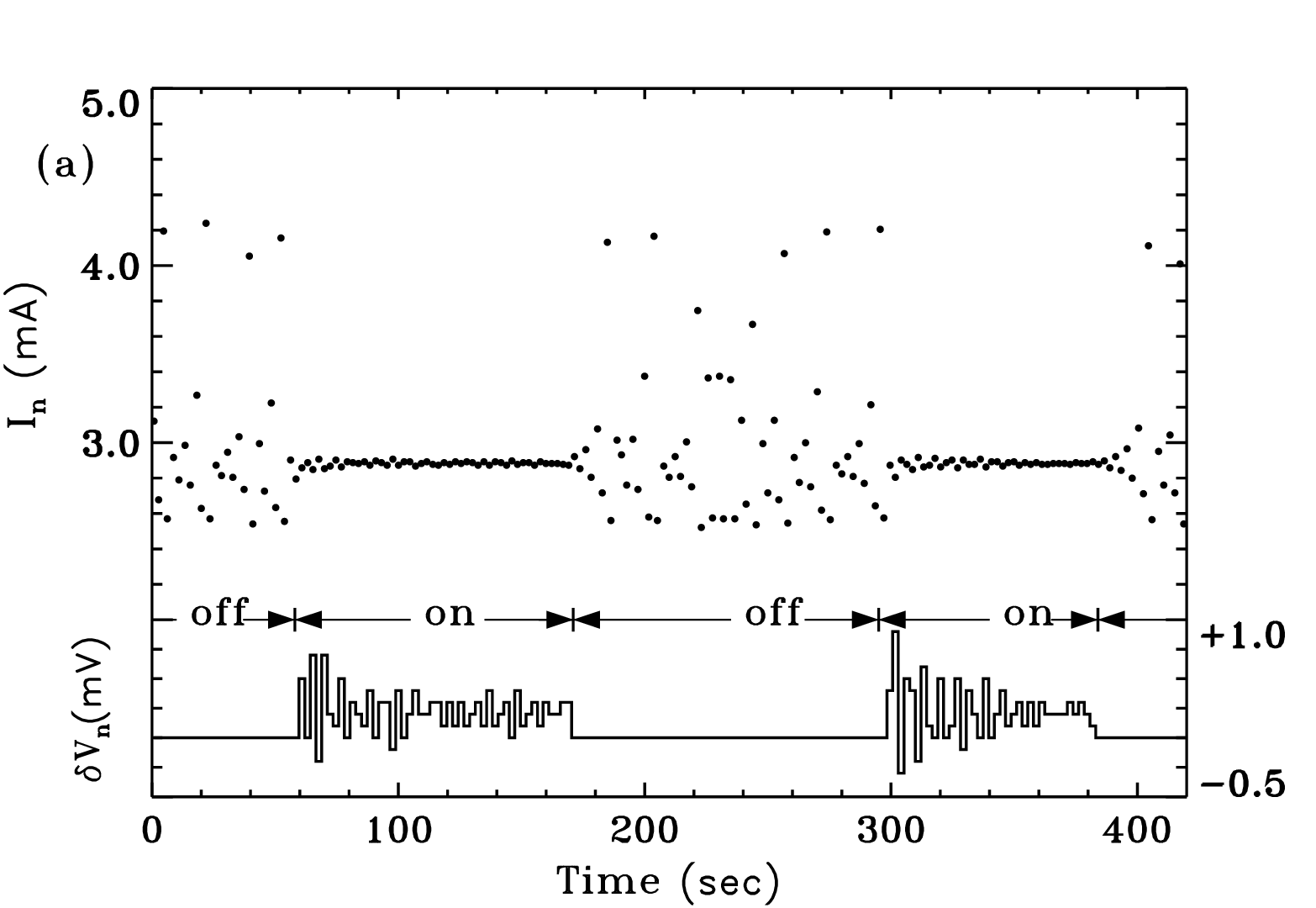}
\vspace{0.1in}
\epsfxsize=4.3in
\epsffile{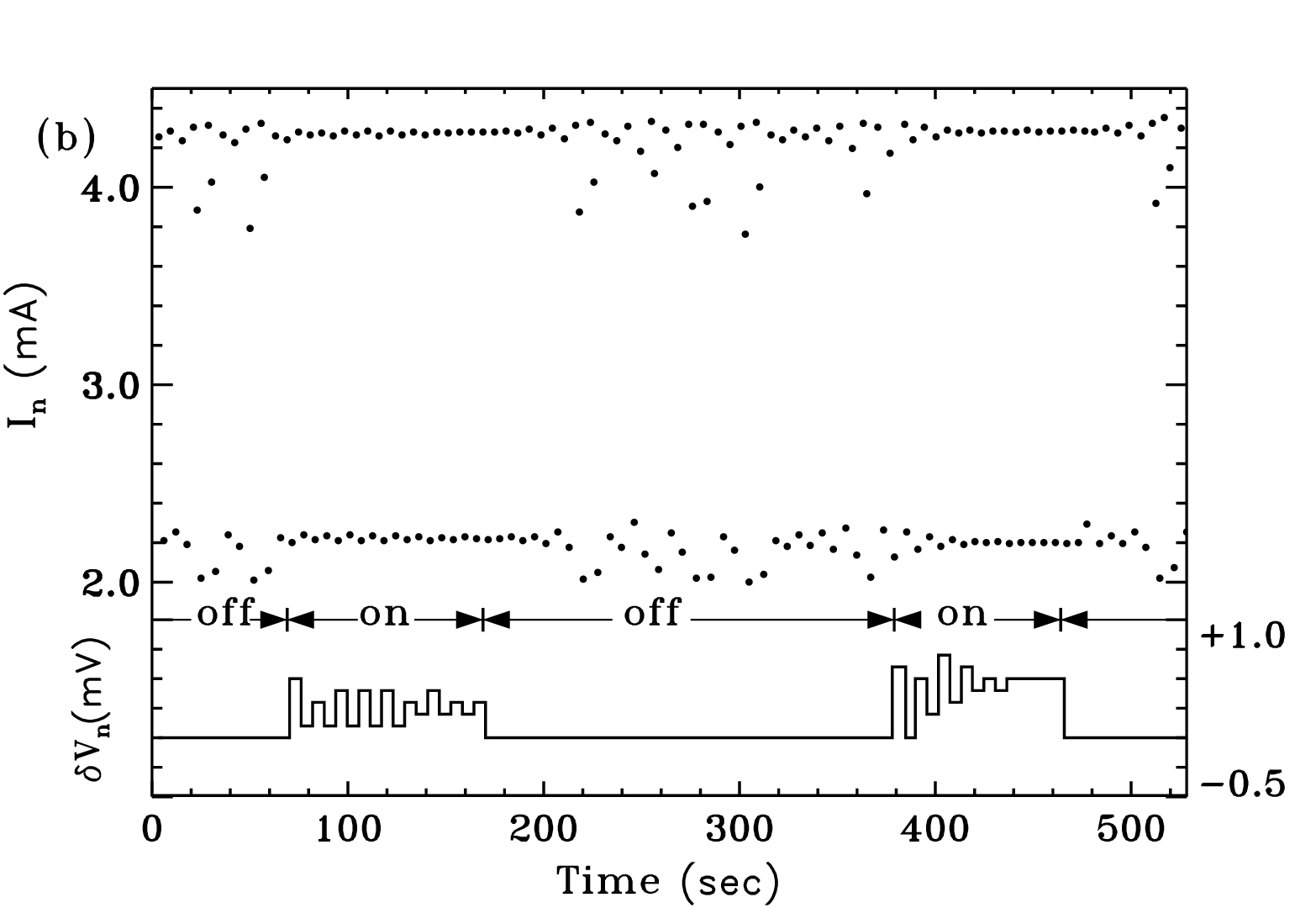}
\end{center}
\begin{center}
\begin{minipage}{30pc}
\caption[]{{\small The minima in anodic current plotted over a time 
period during
which the control algorithm is switched on and off twice.  The perturbations
added to the anodic potential to maintain control are shown in the bottom 
graph. (a) shows period-1 control and (b) shows period-2 control.}
\label{p1p2ctrl}}
\end{minipage}
\end{center}

\end{figure}

Examples of the successful stabilizing of period-1 and period-2
orbits are shown in Fig.~\ref{p1p2ctrl}.  
The rotation rate for the 
copper disk anode was $\omega =$ 2670~rpm, the anodic potential with 
control off was $V_0 =$ 0.724 V, and the values of the RPF proportionality
constants were $K = -$5.0 mV/mA and $R = -$0.3 for the period-1 control 
shown in Fig.~\ref{p1p2ctrl}a.  For the period-2 control shown in
Fig.~\ref{p1p2ctrl}b, $\omega =$ 2160~rpm, $V_0 = $ 0.745 V, $K = -$5.0 mV/mA,
and $R = $0.21.  As described in reference~\ref{bib:RPFexp1}, 
we chose a region of
parameter space where the electrochemical
system exhibits a sequence of periodic mixed mode oscillations separated by
bands of chaotic behavior as a function of anodic potential.
The period-1 control was done while in the chaotic region between a period-1
state with large amplitude oscillations only and a period-2 state with one
large amplitude oscillation followed by one small oscillation.
We were unsuccessful in controlling on the period-2 oscillation in this
chaotic band.  Successful control of a period-2 state was attained
in the chaotic band between the period-2 (one large--one small) and 
period-3 (one large and two small) mixed mode oscillations.  We  
found that control of the period-2 orbit was much more difficult than the
period-1.  This is partially because the period is about twice as long 
(about 5 sec) and the feedback corrections are made just once each cycle.
Thus the system can drift further away from its fixed point before the
next feedback correction is calculated and applied.

\begin{figure}[pt]
\begin{center}
%\vspace{7.0 in}
\leavevmode
\epsfxsize=3.35in
\epsffile{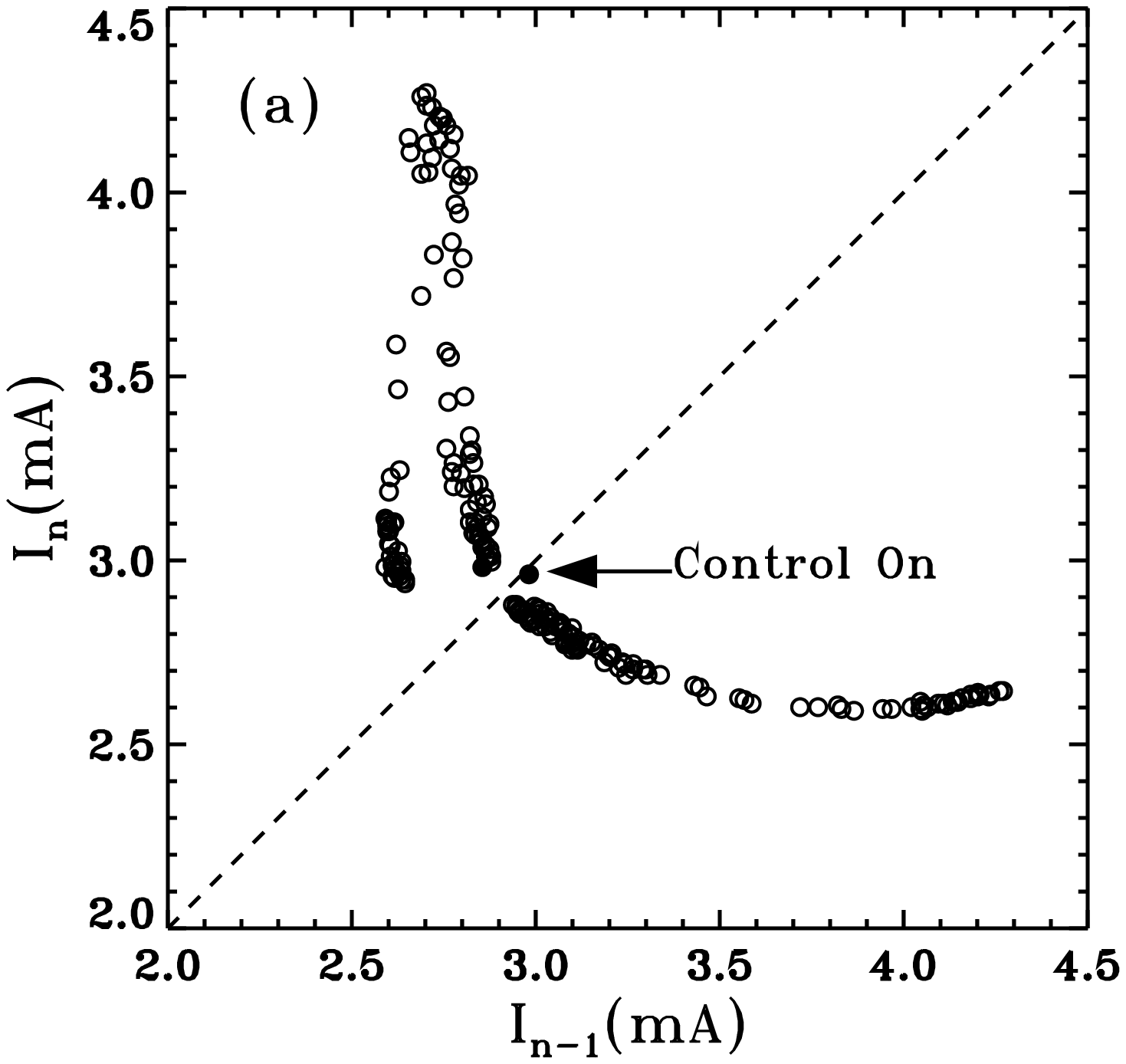}
\vspace{0.1in}
\epsfxsize=3.35in
\epsffile{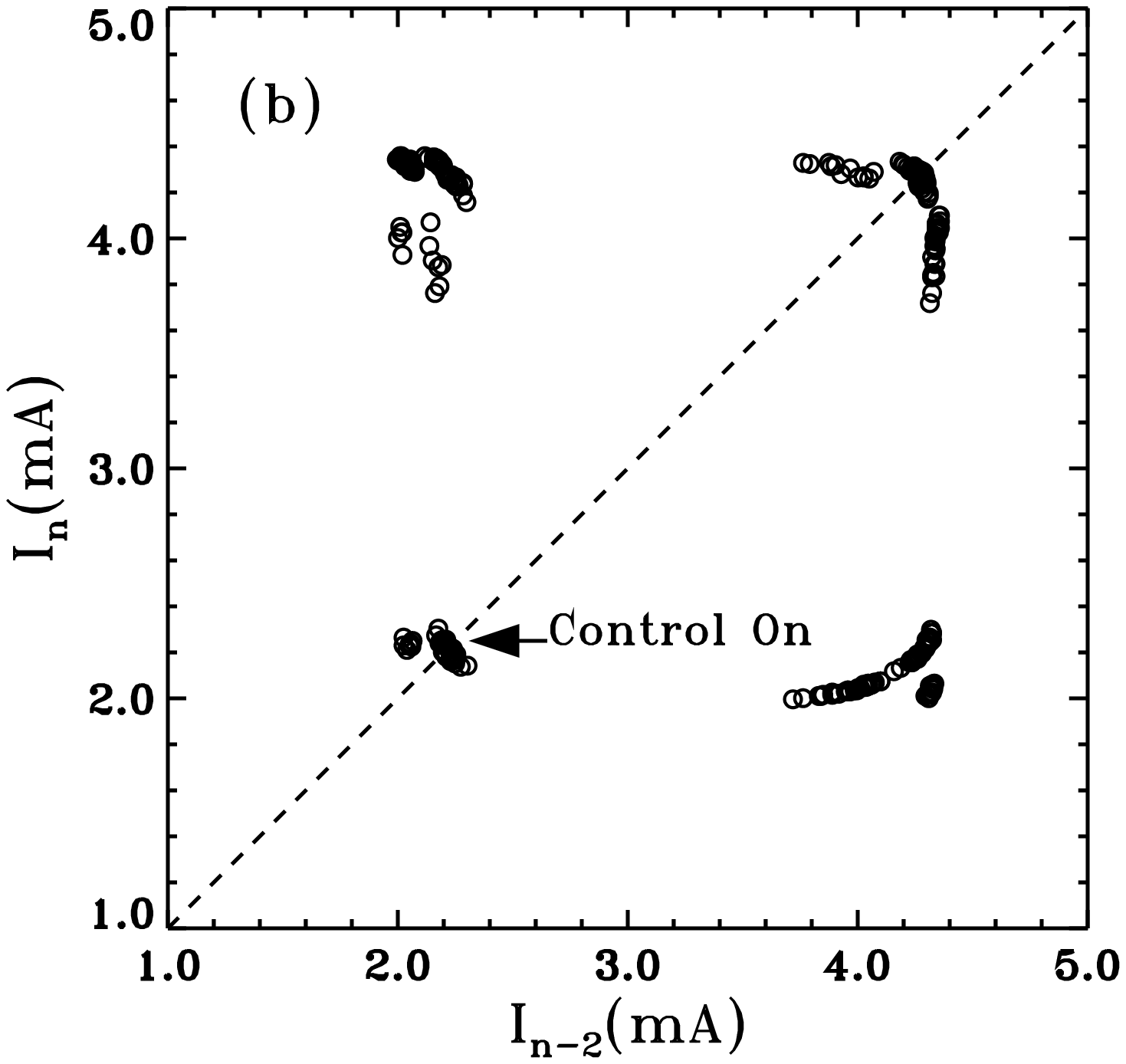}
\end{center}
\begin{center}
\begin{minipage}{30pc}
\caption[]{{\small (a) The first iterate return map (open circles) obtained
from the sequence of current minima
for the chaotic state shown in Fig.~2a.  The superimposed filled circles 
are the minima in the anodic current while the control algorithm was 
implemented.  (b) the corresponding
second-iterate return map for the period-2 case shown in Fig.~2b.}
\label{rtmap}}
\end{minipage}
\end{center}
\end{figure}

\begin{figure}[pt]
\begin{center}
%\vspace{7.0 in}
\leavevmode
\epsfxsize=3.4in
\epsffile{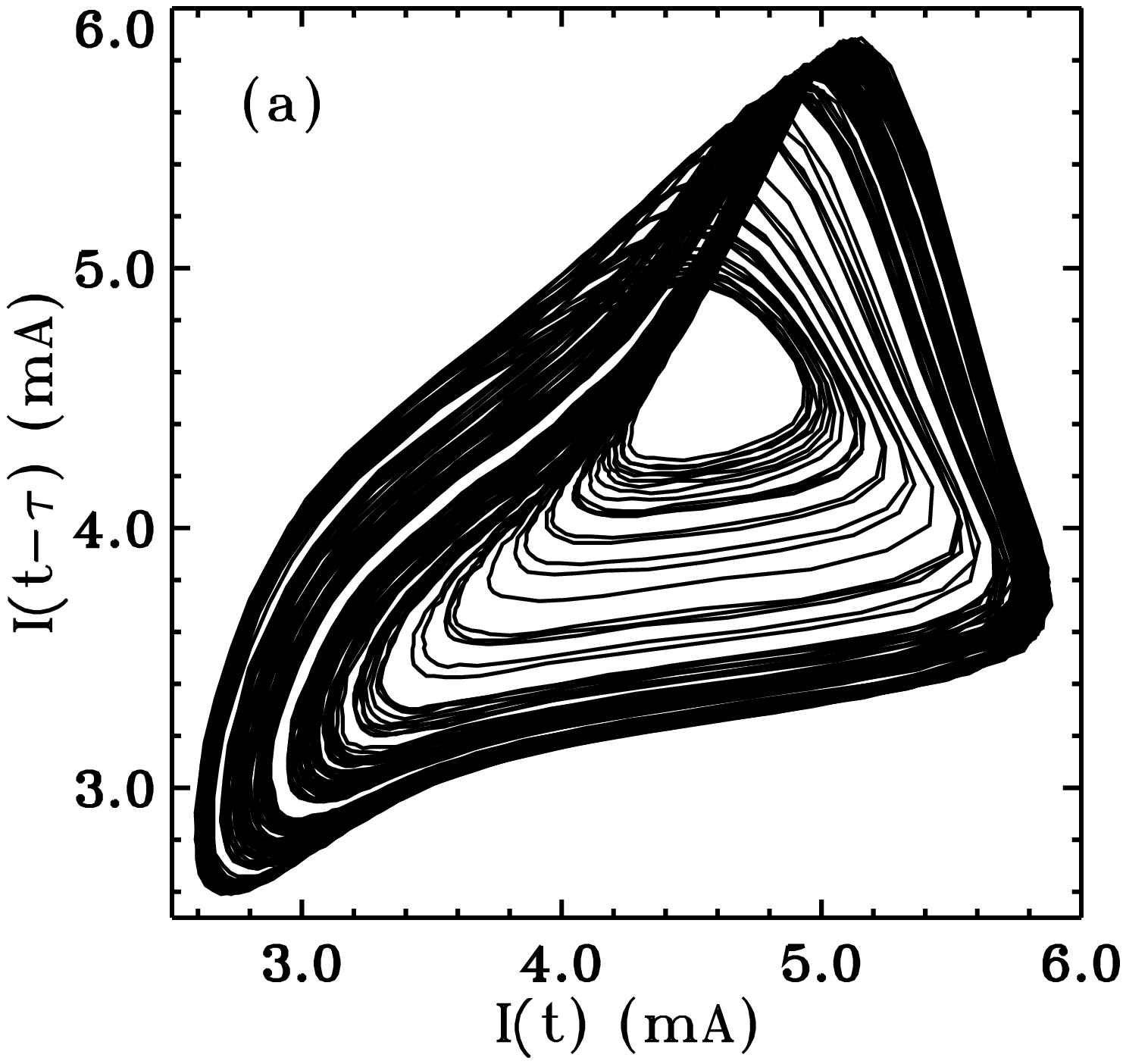}
\vspace{0.1in}
\epsfxsize=3.4in
\epsffile{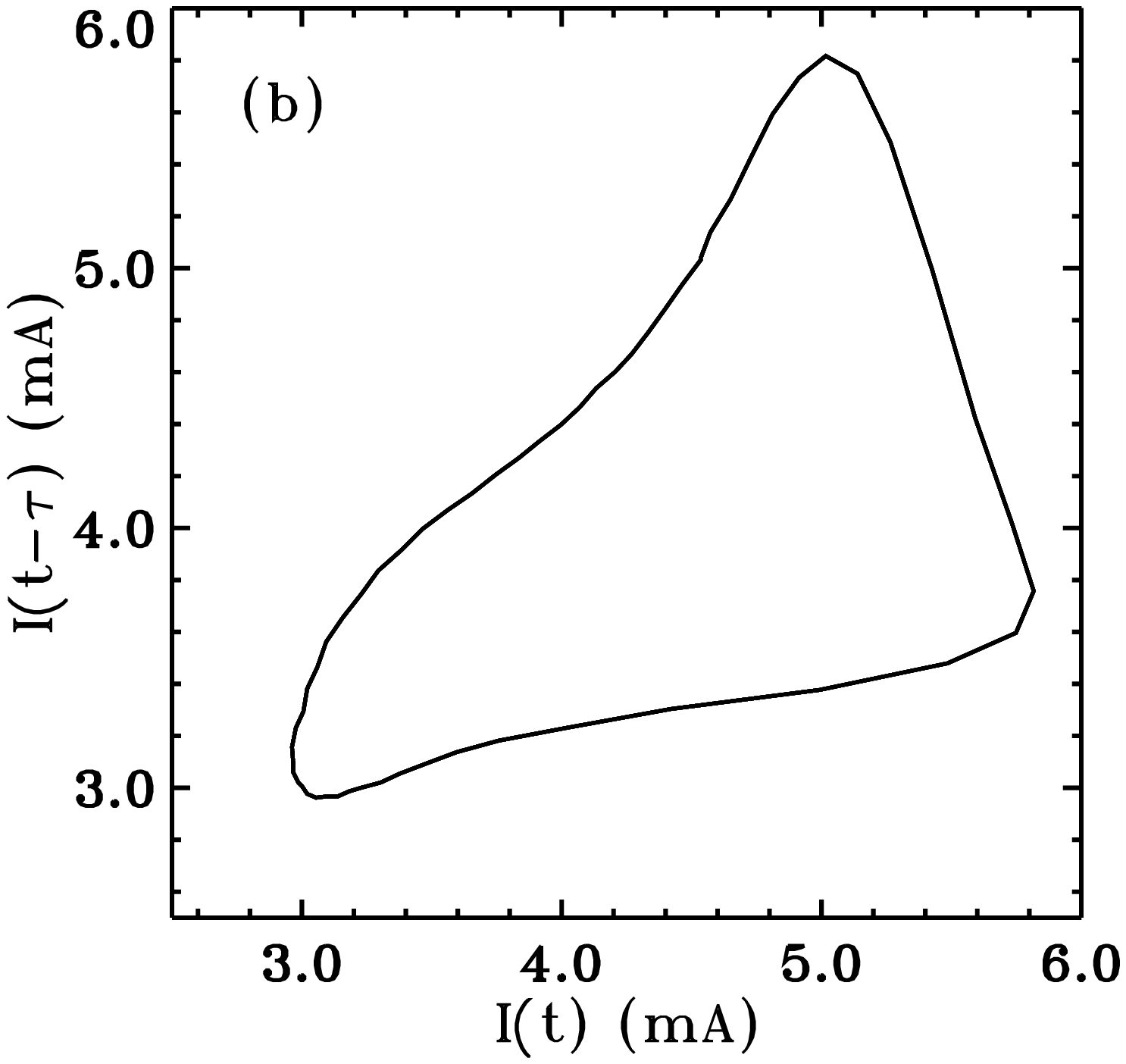}
\end{center}

\begin{center}
\begin{minipage}{30pc}
\caption[]{{\small (a) The two-dimensional time delay embedding 
with $\tau =$ 120 msec
showing the chaotic
attractor for the situation shown in Figs.~2a and 3a. (b) the corresponding
period-1 trajectory with control on.}\label{p1att}}
\end{minipage}
\end{center}
\end{figure}

\begin{figure}[pt]
%\vspace{7.0 in}
\begin{center}
\leavevmode
\epsfxsize=3.4in
\epsffile{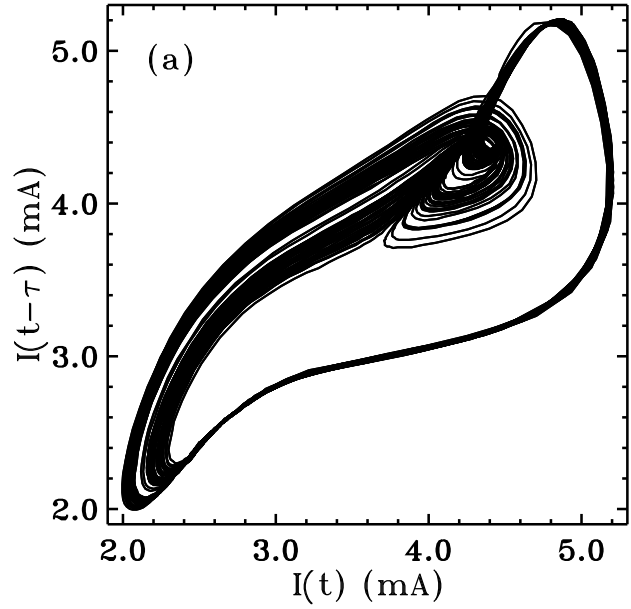}
\vspace{0.1in}
\epsfxsize=3.4in
\epsffile{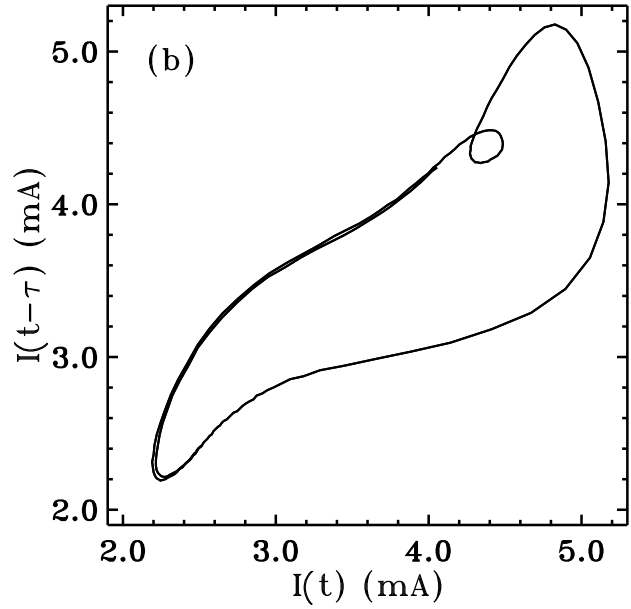}
\end{center}
\begin{center}
\begin{minipage}{30pc}
\caption[]{{\small (a) The two-dimensional time delay embedding 
with $\tau =$ 120 msec
showing the chaotic
attractor for the situation shown in Figs.~2b and 3b. (b) the corresponding
period-2 trajectory with control on.}\label{p2att}}
\end{minipage}
\end{center}

\end{figure}

The corresponding return maps for the two cases are shown in 
Fig.~\ref{rtmap} while Fig.~\ref{p1att} and \ref{p2att} show the 
chaotic attractors and
controlled periodic orbits reconstructed from the time series of the anodic
current using a two-dimensional time delay embedding.

\clearpage

\vspace{14pt}
\noindent
{\em 4.2.~~Why use RPF method}
\vspace{14pt}

  Feedback methods used to control chaos as discussed in this paper
are most closely related to the algorithm first
proposed by Ott, Grebogi, and Yorke (OGY)$^{\ref{bib:OGY}}$.  
The original OGY method required the measurement of two system variables
at the Poincar\'{e} surface (in a three dimensional system) to determine
the appropriate feedback for control. 
However, many experimental situations (such as that described in this paper)
donot easily allow the measurement of more than one variable.
It was recently
found$^{\ref{bib:RPFth1},\ref{bib:DN},\ref{bib:AGOY}}$ that the
original OGY method must be modified 
to apply to the situation where single time series data is used.  In general,
this requires nontrivial modification of the algorithm and makes the prescribed
change in the control parameter on the $n$th cycle depend on the changes 
that were made 
on previous cycles.  It was shown in reference~\ref{bib:RPFth1} that in highly
dissipative systems this reduces to a simple recursive algorithm where the
change on the $n$th cycle depends only on the change made on the $(n-1)$th
cycle as indicated in Eq.~\ref{eq:rpf}.  Furthermore, it was shown that
the recursive term goes to zero ($R = 0$) if the attractor
(in the neighborhood of the fixed point at the Poincar\'{e} section)  
does not shift in the direction normal to its plane in state space
when small changes are made in the control parameter. 

\begin{figure}[hbt]
%\vspace{3.7 in}
\begin{center}
\leavevmode
\epsfxsize=4.2in
\epsffile{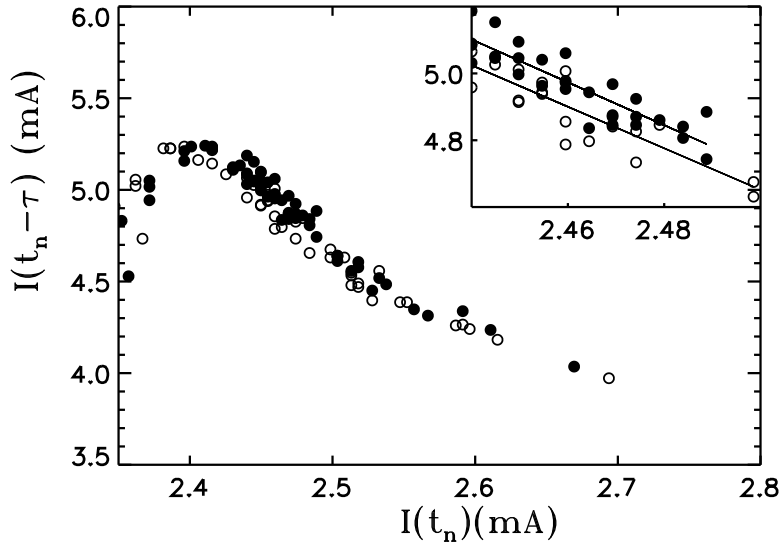}
\end{center}
\begin{center}
\begin{minipage}{30pc}
\caption[]{{\small Poincar\'{e} sections reconstructed using time delay
embedding of the measured time series of the anodic current.
The anodic current is minimum at times $t_n$ and 
$\tau = $~600~msec.  The open
and closed circles are for values of the anodic potential that
differ by 4~mV.}\label{shift}}
\end{minipage}
\end{center}    

\end{figure}

Figure~\ref{shift} shows the Poincar\'{e} section of the reconstructed 
attractor in a delay
coordinate embedding for the electrochemical cell at parameter values 
where we used RPF to control on a period-1 orbit.  The time $t_n$ is
the time when the anodic current goes through the $n$-th minimum near the 
period-1 orbit.  Two sets of data are shown with the anodic 
potential, $V$, held constant in each case.  The open circles are for
$V =$ 0.720~V and the closed circles for $V =$ 0.724~V.  This figure
showing experimental data should be compared with Fig.~1b of 
reference~\ref{bib:RPFth1}.  While the observed experimental shift in the
Poincar\'{e} section is small, there is clear evidence that a shift is
present.

\vspace{14pt}
\noindent
{\em 4.3.~~Theoretical robustness of RPF method}  
\vspace{14pt}

   We have found in our experimental work that there is often a
rather large
range in the values of the RPF proportionality constants $K$ and $R$ that
will still be effective in controlling the system.  This is fortunate since
it allows us to make estimates of $K$ and $R$ from just a few experimental
points in the neighborhood of the desired fixed point during the precontrol
phase of the experiment.  Here, we briefly derive the RPF equations 
for $K$ and
$R$ using methods of control theory$^{\ref{bib:Ogata}, \ref{bib:RGOD}}$  
in order to to shed light on the theoretical robustness of the RPF method.

   We start with the general 1D return-map equation (Eq.~2 from 
reference~\ref{bib:RPFth1})
\begin{equation}
X_{n+1} = f(X_n;p_{n-1}, p_n), \label{eq:R1-dmap}
\end{equation}
where the notation is that used in reference~\ref{bib:RPFth1},
$X_n$ is the value of the measured variable at the start of the
$n$th Poincar\'{e} cycle and $p_n$ is the value of the control 
parameter during
the $n$th cycle.  For control on a period-1 orbit, we are interested
in the natural dynamics of the system described by
Eq.~\ref{eq:R1-dmap} near the fixed point, $X_F$, of the period-1 orbit for
$p = p_0$; $X_F=f(X_F,p_0,p_0)$.  To first 
order in $\delta X_n=(X_n-X_F)$, $\delta p_{n-1}$, and 
$\delta p_n$, we have
\begin{equation}
\delta X_{n+1} = \mu \delta X_n + w \delta p_{n-1} + v \delta p_n 
\label{eq:Rmap},
\end{equation}
where $\mu = \partial f(X_F,p_0,p_0)/\partial X_n$,
$w = \partial f(X_F,p_0,p_0)/\partial p_{n-1}$, and 
 $v= \partial f/\partial p_n$.

 In reference~\ref{bib:RPFth1}, the RPF algorithm is obtained by using 
the optimum possible strategy of choosing 
$\delta p_n$ such that the system is 
brought to the fixed point as quickly as possible, namely two
Poincar\'{e} cycles.    
Taking   
$\delta X_{n+2}=0$ and
$\delta p_{n+1}=0$, then the first and second iterate of 
Eq.\ (\ref{eq:Rmap}) gives the recursive control algorithm Eq.~\ref{eq:rpf}
with $R$ and $K$ given by 
\begin{equation}
K = -\frac{\mu^2}{(\mu v + w)},
\mbox{  and  } R = -\frac{\mu w}{(\mu v + w)}.
\label{eq:KR1}
\end{equation}
As shown in reference~\ref{bib:RPFth1}, $w=(1-\mu)g_b$ and $v=(1-\mu)g_u$, 
and Eqs.~\ref{eq:KR2} are equivalent to Eqs.~\ref{eq:KR1}.

Here we take a different approach.  We {\em assume} there is a bilinear
relationship between $\delta p_{n}$ and $(\delta X_n,\delta p_{n-1})$
\begin{eqnarray}
\delta  p_n =  H\,\delta X_n + G\,\delta p_{n-1},
\label{eq:rpf2}
\end{eqnarray}
where $H$ and $G$ are constants yet to be determined.
Equations~\ref{eq:Rmap} and \ref{eq:rpf2} form a two-dimensional discrete
map describing the full control system including both the dynamics of the 
nonlinear system and the recursive feedback control strategy.  
The two-dimensional map is put in more conventional 
form if we shift the $n$ labels on
the $p_n$ by 1.  This can be accomplished by defining a new parameter,
$\delta \hat{p}_n = \delta p_{n-1}$.  Substitution into Eqs.~\ref{eq:Rmap} and
\ref{eq:rpf2} gives
\begin{eqnarray}
\delta X_{n+1}& = &\mu\,\delta X_n + w\,\delta \hat{p}_n 
                             + v\, \delta \hat{p}_{n+1}\label{eq:a}\\
\delta \hat{p}_{n+1}& = &H\, \delta X_n + G\,\delta \hat{p}_n. \label{eq:b}
\end{eqnarray}  
The conventional 2D map is formed by substituting Eq.~\ref{eq:b} into the
last term of Eq.~\ref{eq:a} giving
\begin{equation}
\left [{\begin{array}{c} \delta X_{n+1} \\ \delta \hat{p}_{n+1} \end{array} }
\right ]  = \tilde{M} 
  \left [{\begin{array}{c} \delta X_{n} \\ \delta \hat{p}_{n} \end{array} }
   \right ],
\label{eq:2dmap}
\end{equation}
where 
\begin{equation}
\tilde{M} = \left [
             { \begin{array}{cc}
                \mu+vH & w+vG \\
                  H    &   G 
               \end{array} }
            \right ].
\label{eq:M}
\end{equation}

The full control system defined by Eq.~\ref{eq:2dmap} has a fixed 
point at $\delta X_n = \delta \hat{p}_n = 0$.  The control system will produce
the desired control if this fixed point is stable. 
The parameters $\mu$, $v$, and $w$ are determined by the 
dynamics of the system near the fixed point and we are free to adjust
$H$ and $G$.  For successful feedback
control, the values of $H$ and $G$ must be chosen so that the fixed
point of Eq.~\ref{eq:2dmap} is stable. 

The stability of the fixed point is determined by the magnitude of the
eigenvalues of $\tilde{M}$:
\begin{equation}
 \lambda_{1,2} = \frac{-\mbox{Tr}\tilde{M} \pm \sqrt{(\mbox{Tr}\tilde{M})^2
        -4\mbox{Det}\tilde{M}}}{2},
\label{eq:s1}
\end{equation}
where Tr$\tilde{M} = \mu + vH + G$ and Det$\tilde{M} = \mu G-wH$.
The fixed point will be stable if 
\begin{equation}
|\lambda_1|<1 \mbox{ and } |\lambda_2| < 1.
\label{eq:s2}
\end{equation}
Equations~\ref{eq:s1} and \ref{eq:s2} put conditions on $H$ and $G$ such
that the control strategy will work.

The best values of $H$ and $G$ would make $\lambda_1= \lambda_2=0$ and
the fixed point of the control system becomes superstable.  The superstable
condition is satisfied if $H=H^*$ and $G=G^*$ where  
\begin{eqnarray}
\mbox{Tr}\tilde{M}& = 0 = &\mu + vH^* + G^* \label{eq:ss1} \\
\mbox{Det}\tilde{M}& = 0 = &\mu G^* - wH^*.
\label{eq:ss2}
\end{eqnarray}
The values of $H^*$ and $G^*$ that satisfy Eqs.~\ref{eq:ss1} and \ref{eq:ss2}
are
\begin{equation}
H^* = -\frac{\mu^2}{(\mu v + w)},
\mbox{  and  } G^* = -\frac{\mu w}{(\mu v + w)}.
\label{eq:HG1}
\end{equation}

Equations~\ref{eq:HG1} are identical to Eqs.~\ref{eq:KR1} for the 
RPF algorithm with $K \equiv H^*$ and $R \equiv G^*$ and we have shown
that the RPF algorithm forms a superstable control system.  Of course, it
is not necessary to have the optimum values for $K$ and $R$ for the control
to be successful and there is a range of values satisfying the
stability conditions of Eqs.~\ref{eq:s1} and \ref{eq:s2}. 

Finally, if we substitute $H=H^*$ and $G=G^*$ into $\tilde{M}$, then
the map describing the superstable control system becomes
\begin{eqnarray}
\delta X_{n+1} & = & \frac{1}{(\mu+w)}
                        (\mu \delta X_n + w^2 \delta \hat{p}_n)\\
\delta \hat{p}_{n+1} & = & \frac{1}{(\mu+w)}
                        (-\mu^2 \delta X_n - \mu v \delta \hat{p}_n).
\end{eqnarray}
Iterating these equations once shows explicitly that, for the superstable
RPF control conditions, $\delta X_{n+2} \equiv 0$ and 
$\delta \hat{p}_{n+2} \equiv 0$ for {\em any} starting values of
$(\delta X_n, \delta \hat{p}_n)$.  
Of course, $\delta X_n$ and $\delta \hat{p}_n$ must be
small enough so that the linearization of the dynamics about $X_F$ is valid. 

\vspace{14pt}
\noindent
{\bf 5.~~Acknowledgements}
\vspace{14pt}

   We thank Alan Markworth at 
Battelle, Columbus for his continued thoughtful
comments and suggestions on this project.  We have also benefited from 
discussions with Don Weekley, Gregg Johnson, Andreas Rhode, 
Markus L\"{o}cher, and Earle Hunt 
at Ohio University regarding the experiment and the subtleties of
nonlinear dynamics and chaos control. This work was 
supported in part by a Battelle subcontract of the Electric Power Research
Institute (EPRI) research project contract RP2426-25 and by Ohio University
Research Challenge Grant RC89-107.

\vspace{14pt}
\noindent
{\bf 6.~~References}

\begin{enumerate}
\item \label{bib:BZctrl1}
V.\ Petrov, V.\ G\'{a}sp\'{a}r, J.\ Masere, and K.\ Showalter,
{\em Nature}, {\bf 361} (1993) 240.

\item \label{bib:RPFexp1} 
P.\ Parmananda, P.\ Sherard, R.\ W.\ Rollins, and H.\ D.\
Dewald, {\em Phys.\ Rev.} {\bf E47}, (1993) R3003.

\item \label{bib:RPFth1} 
R.\ W.\ Rollins, P.\ Parmananda, and P.\ Sherard, 
{\em Phys.\ Rev.} {\bf E47}, (1993) R780.

\item \label{bib:PPS2} 
V.\ Petrov, B.\ Peng, and K.\ Showalter, {\em J.\ Chem.\ Phys.}
{\bf 96}, (1992) 7506. 

\item \label{bib:EH} 
E.\ R.\ Hunt, {\em Phys.\ Rev.\ Lett.} {\bf 67}, (1991) 1953.

\item \label{bib:Ogata}
K.\ Ogata, {\em Control engineering}, Second Ed.\ 
(Prentice-Hall, Englewood Cliffs, NJ, 1990).

\item \label{bib:RGOD}
F.\ J.\ Romeiras, C.\ Grebogi, E.\ Ott, and W.\ P.\ Dayawansa,
{\em Physica} {\bf D 58}, (1992) 165.

\item \label{bib:DPR1} 
H.\ D.\ Dewald, P.\ Parmananda, and R.\ W.\ Rollins, 
{\em J.\ Electroanal.\ Chem.} {\bf 306}, (1991) 297. 

\item \label{bib:DPR2} 
H.\ D.\ Dewald, P.\ Parmananda, and R.\ W.\ Rollins, 
{\em J.\ Electrochem.\ Soc.} {\bf 140}, (1993) 1969.

\item \label{bib:OGY} 
E.\ Ott, C.\ Grebogi, and J.\ A.\ Yorke, 
{\em Phys.\ Rev.\ Lett.} {\bf 64}, (1990) 1196.

\item \label{bib:DN} 
U.\ Dressler and G.\ Nitsche, {\em Phys.\ Rev.\ Lett.} 
{\bf 68}, (1992) 1.

\item \label{bib:AGOY}
D.\ Auerbach, C.\ Grebogi, E.\ Ott, and J.\ Yorke, {\em Phys.\ Rev.\ Lett.}
{\bf 69}, (1992) 3479.

\end{enumerate}

\end{document}